\documentclass[twocolumn,amssymb,amsfonts,
prb,aps,floats,groupedaddress]{revtex4}  

\usepackage[final]{epsfig}

\begin{document}
\draft

\title[]{Theory of Spin-Resolved Auger-Electron
  Spectroscopy from Ferromagnetic 3d-Transition Metals}

\author{T. Wegner}
\email[]{torsten.wegner@physik.hu-berlin.de}
\author{M. Potthoff}
\author{W. Nolting}
\affiliation{Theoretische Festk\"orperphysik, Institut f\"ur Physik,
  Humboldt-Universit\"at zu Berlin, Invalidenstra{\ss}e 110, 10115
  Berlin, Germany}

\begin{abstract}
  CVV Auger electron spectra are calculated for a multi-band Hubbard
  model including correlations among the valence electrons as well as
  correlations between core and valence electrons.  The interest is
  focused on the ferromagnetic 3d-transition metals.  The Auger line
  shape is calculated from a three-particle Green function.  A
  realistic one-particle input is taken from tight-binding
  band-structure calculations.  Within a diagrammatic approach we can
  distinguish between the \textit{direct} correlations among those
  electrons participating in the Auger process and the
  \textit{indirect} correlations in the rest system. The indirect
  correlations are treated within second-order perturbation theory for
  the self-energy. The direct correlations are treated using the
  valence-valence ladder approximation and the first-order
  perturbation theory with respect to valence-valence and core-valence
  interactions. The theory is evaluated numerically for ferromagnetic
  Ni. We discuss the spin-resolved quasi-particle band structure and
  the Auger spectra and investigate the influence of the core hole.
\end{abstract}
\pacs{79.20.Fv,71.20.Be,75.60.Ej}
\maketitle

\section{Introduction}
\label{sec:intro}
Auger-electron spectroscopy (AES) and the complementary
appearance-potential spectroscopy (APS) have become valuable tools for
investigating the electronic structure of solids and solid
surfaces~\cite{Fug81,WM81,AH83,Wei84,WAC88,SCSG89,Ram91,SRC93}. They
represent highly element specific and non-destructive methods with a
comparatively simple experimental set up. The Auger line shape from a
core-valence-valence (CVV) process yields information on the occupied
part of the valence band, while the APS provides insight into the
unoccupied valence states.  However, much effort has been spent on the
detailed interpretation of the spectra.
\\
Lander~\cite{Lan53} suggested that the spectrum obtained by AES (APS) is
given as the self-convolution of the occupied (unoccupied) valence
density of states (DOS).  On the other hand, Powell~\cite{Pow73}
discovered the CVV Auger line shape of Ag to behave ``anomalously'' in
the sense of Lander's self-convolution model. These anomalous features
are by now well known to be caused by correlation effects dominating the
electronic properties of various solids. Therefore, AES (APS) seems to
be a useful technique to study electron-correlation effects, but it is
doubtful whether it is able to compete with one-particle spectroscopies,
such as photoemission (inverse photoemission), in deriving the DOS by
deconvolution.
\\
In the theoretical treatment of the CVV Auger process, there are mainly
two problems.  The first one is to take into account the correlation
effects.  Here one may distinguish between the \textit{direct} and
\textit{indirect} correlations. The direct correlations describe the
correlations of those electrons which participate in the Auger process.
They are responsible for the most prominent effects in the Auger line
shape as compared to the self-convolution.  On the other hand, the
indirect correlations among the electrons in the rest system manifest
themselves in the quasi-particle density of states (QDOS) as a
renormalisation of the one-particle DOS.
\\
The second problem is the calculation of the transition-matrix elements
for the Auger process as well as the scattering of the outgoing Auger
electron (cf.\ Refs.~\onlinecite{HWMR88,KR97}).  These effects will
(slightly) modify the bare line shape and may become important for a
refined interpretation of experimental data.  Within the present paper,
however, we set aside this second problem and like to concentrate on
electron-correlation effects in AES from ferromagnetic 3d-transition
metals.
\\
Within the framework of the single-band Hubbard model, correlation
effects can be treated exactly for systems with completely filled or
empty bands, as was first shown by Cini and
Sawatzky~\cite{Cin77,Saw77,SL80}. The generalisation to the case of
degenerate bands was introduced in Ref.~\onlinecite{Cin78} and further
analysed in Ref.~\onlinecite{NGE92}, for example. These results may also
be extended to include the core-valence interaction~\cite{PBNB93}.
\\
Considering the more general case of partially filled bands introduces
several complications concerning the indirect as well as the direct
correlations.  For the indirect valence-valence correlations there is a
number of approximation schemes applicable to a multi-band Hubbard
model. A method which reproduces the experimentally observed Curie
temperature quite well, especially for Ni, is the spectral density
approach~\cite{NBDF89,NVF95}.  Other approaches are, for example, the
generalisation of the single-band modified perturbation
theory~\cite{PWN97} to the multi-band model~\cite{KK97,LK98} and
Quantum-Monte-Carlo simulations~\cite{HV98} in connection with the
dynamical mean-field theory~\cite{MV89}.  However, these methods suffer
from some necessary restrictions concerning the completeness of the
Coulomb-matrix. This is not the case for the fluctuation
exchange~\cite{KL99} and the Hubbard~I approximation~\cite{LK98}, for
example. For a more detailed discussion on the indirect valence-valence
correlations see Ref.~\onlinecite{LK98}.
\\
For the treatment of the direct correlations, one has the
exact-diagonalisation method~\cite{SM98} for small systems and the
equation-of-motion method~\cite{Drc89,KD92} with its in general
uncontrolled termination of the hierarchy of the equations of motion.
Another approximate solution is the valence-valence ladder (VV ladder)
approximation~\cite{Cin79,TDDS81,DK84,Nol90,NGE91} and its
generalisation to include the core-valence
interaction~\cite{PBBN93,PBB94,PBNB94,PBBN95} (CVV ladder).  In
particular one has to account for the broken translational symmetry in
the initial state of AES, caused by the presence of the core hole and
its screening due to the valence electrons.  In the final state this
interaction is responsible for the sudden response of the valence
electrons due to the destruction of the core hole.  In the limit of
completely filled or empty bands the ladder approximations recover the
above-mentioned exact solution.

Here the interaction strength is taken as the control parameter, which
is correct in the weak-coupling regime. We are aware that this method
has restrictions for the 3d-transition metals. However, in this work we
prefer a common treatment of one-particle spectroscopies (photoemission
and inverse photoemission) and two-particle spectroscopies like AES. To
be concrete, we will use the second-order perturbation theory around the
Hartree-Fock solution~\cite{AGD,KM81,SAS92} for the indirect
valence-valence correlations.  The direct correlations will be treated
by applying two different methods, i.\,e.\ the VV ladder approximation
and the first-order perturbation theory in the valence-valence and
core-valence interaction.
\\
Within this approach it is possible to include a realistic one-particle
input taken from tight-binding band-structure calculations.  We do not
only account for the degeneracy of the 3d-band but also for the
hybridisation with the 4s and 4p states. The theory is formulated and
evaluated for a non-orthogonal basis set where the states can be
distinguished by the angular momentum quantum number and the cubic
harmonic index. This facilitates the interpretation of the resulting
spectra. Furthermore, we do not restrict ourselves to correlations among
the final-state holes only and include core-hole effects from the very
beginning. This implies the necessity for a proper treatment of the
initial state where the core-hole screening breaks the translational
symmetry.  The theory is implemented numerically and evaluated for
ferromagnetic Ni.

The paper is organised as follows.  In the next section we will
introduce the model under consideration. In section~\ref{sec:auger} we
give the expression for the Auger intensity.  Section~\ref{sec:indirect}
concentrates on the indirect and section~\ref{sec:direct} on the direct
correlations. Finally, section~\ref{sec:sum} concludes the paper. Some
details concerning the non-orthogonal basis set are given in the
appendix~\ref{sec:nobs}.
%%%%%%%%%%%%%%%%%%%%%%%%%%%%%%%%%%%%%%%%%%%%%%%%%%%%%%%%%%%%%%%%%%%%%
\section{Model}
\label{sec:model}
The Hamiltonian $\mathcal{H}=H_0-\mu N+H_{\rm I}$ is decomposed into a
one-particle part $H_0-\mu N$ and an interaction part $H_{\rm I}$. $N$
is the operator for the particle number. The one-particle part describes
non-interacting valence and core electrons:
\begin{eqnarray}
  \label{eq:h0}
  H_0 - \mu N
  &=& \sum_{i,i^\prime,\sigma, \atop L,L^{\prime}}
  \left(t_{ii^\prime}^{LL^{\prime}}-\mu S_{ii^\prime}^{LL^{\prime}}\right)
  \,c_{iL\sigma}^{\dag}c_{i^\prime L^{\prime}\sigma}
  \nonumber\\&&+
  \sum_{i,\sigma}\left(\epsilon_c-\mu\right)
  \,b_{i\sigma}^{\dag}b_{i\sigma}
\end{eqnarray}
The index $i$ refers to lattice sites, $\sigma$ is the spin-index
($\sigma=\uparrow,\downarrow$), $L=\{l,m\}$ is the orbital index with
angular momentum quantum number $l$ and cubic harmonic index $m$.
$c_{iL\sigma}^{\dag}$ ($c_{iL\sigma}$) denotes the creation
(annihilation) operator of a valence electron at the lattice site $i$
with spin $\sigma$ and orbital index $L$ while $b_{i\sigma}^{\dag}$
($b_{i\sigma}$) creates (annihilates) a core electron. The core states
are assumed to be non-degenerate and dispersionless with the
one-particle energy $\epsilon_c$ well below the chemical potential
$\mu$. The hopping integrals $t_{ii^\prime}^{LL^{\prime}}$
\begin{equation}
  \label{eq:hopping}
  \langle iL\sigma|h^{\rm BS}|i^\prime L^\prime \sigma^\prime \rangle =
  t_{ii^\prime}^{LL^{\prime}}\delta_{\sigma\sigma^\prime}
\end{equation}
($h^{\rm BS}$ denotes the Hamiltonian of the tight-binding
band-structure calculation) are taken from Ref.~\onlinecite{Pap86} as
well as the overlap integrals $S_{ii^\prime}^{LL^{\prime}}$
\begin{equation} 
  \label{eq:overlap}
  \langle iL\sigma|i^\prime L^\prime \sigma^\prime \rangle = 
  S_{ii^\prime}^{LL^{\prime}}\delta_{\sigma\sigma^\prime}\,.
\end{equation}
$t_{ii^\prime}^{LL^{\prime}}$ and $S_{ii^\prime}^{LL^{\prime}}$ refer to
a non-orthogonal basis set (see appendix~\ref{sec:nobs}).  Contrary to
an orthonormal basis set (where the overlap matrix is replaced by
$\delta_{ii^\prime}\delta_{LL^{\prime}}$), the basis states under
consideration can be characterised by the orbital index $L=\{l,m\}$. The
construction operators likewise refer to the non-orthogonal basis and
satisfy the following anti-commutation rules:
\begin{eqnarray}
  \left[c_{iL\sigma},c_{i^\prime L^\prime\sigma^\prime}\right]_+
  &=& 0\nonumber\\
  \left[c_{iL\sigma},c^\dag_{i^\prime L^\prime\sigma^\prime}\right]_+
  &=&\left(S^{-1}\right)_{ii^\prime}^{LL^{\prime}}
  \delta_{\sigma\sigma^\prime}\,.
  \label{eq:acr}
\end{eqnarray}
It should be noted that the action of the creation operator on the
vacuum state $c^\dag_{iL\sigma}|0\rangle$, in general, does not yield
$|iL\sigma\rangle$ (see Eq.~(\ref{eq:constr}) of the
appendix~\ref{sec:nobs}).

To describe the correlations among the valence electrons (VV) as well as
the correlations between valence and core electrons (CV) the interaction
consists of two parts
\begin{eqnarray}
  \label{eq:hi}
  H_{\rm I}
  &= &\frac{1}{2}\sum_{i,\sigma,\sigma^\prime,
    \atop L_1, \ldots ,L_4} U_{L_1 L_2 L_4 L_3}
  \,c_{iL_1\sigma}^\dag c_{iL_2\sigma^\prime}^\dag c_{iL_3 \sigma^\prime}
  c_{iL_4 \sigma}-H_{\rm dc}^{\rm VV}\nonumber\\
  & & +\,\sum_{i,\sigma,\sigma^\prime,L}
  U_{L}^c\,
  n_{iL\sigma} n_{i\sigma^\prime}^c-H_{\rm dc}^{\rm CV}\,.
\end{eqnarray}
Here the occupation number operator for valence electrons is
$n_{iL\sigma}=c_{iL\sigma}^{\dag}c_{iL\sigma}$ and for core electrons
$n_{i\sigma}^c=b_{i\sigma}^{\dag}b_{i\sigma}$. Assuming a strong
screening of the Coulomb interaction, the interaction part is taken to
be purely local.  $U_{L_1 L_2 L_4 L_3}$ are the on-site Coulomb-matrix
elements for the valence electrons.
\\
The electronic structure of the 3d-transition metals may be understood
considering mainly two types of electronic orbitals: The 4s- and
4p-states which form broad free-electron like bands. They should be well
described by the band-structure calculation.  The other group are the
well localised 3d-states which in the solid form relatively narrow bands
positioned around the Fermi energy.  The localised nature of the
3d-electrons gives rise to important dynamic 3d-3d correlation effects
which are believed to be responsible e.\,g.\ for the magnetic behaviour
of the 3d-transition metals. These correlations may not be adequately
taken into account within a mean-field picture.  We thus treat them
separately.
\\
Exploiting atomic symmetries, one is able to express all remaining
Coulomb-matrix elements for the 3d-electrons in terms of three effective
Slater integrals~\cite{STK70,AAL97} ($F^0$, $F^2$, $F^4$) only.  These
integrals are connected to averaged values for direct
\begin{equation}
  U=\frac{1}{25}\sum_{L,L^\prime}U_{LL^\prime LL^\prime}=F^0
\end{equation}
and exchange interaction terms
\begin{equation}
  J=\frac{1}{20}\sum_{L \neq L^\prime}
  U_{LL^\prime L^\prime L}=\frac{F^2+F^4}{14}\,.
\end{equation}
For 3d-elements one has to a good accuracy the ratio $F^2/F^4 \approx
0.625$~\cite{AAL97} to be that of free ions~\cite{STK70}. $U$ and $J$
are treated as free parameters to be fixed by comparison with
experimental results (see section~\ref{sec:indirect}).
\\
The CV interaction part is necessary to describe the core hole effects
in AES.  $U_{L}^c$ in equation~(\ref{eq:hi}) are the Coulomb-matrix
elements between the valence and the core electrons which can be fixed
by assuming complete screening of the core hole by the valence electrons
(see section~\ref{sec:indirect}).
\\
To avoid a double counting of interactions, we subtract the correction
$H_{\rm dc}^{\rm VV(CV)}$ which is to a good approximation the
Hartree-Fock part of the respective interaction term~\cite{AAL97}.
%%%%%%%%%%%%%%%%%%%%%%%%%%%%%%%%%%%%%%%%%%%%%%%%%%%%%%%%%%%%%%%%%%%%%%%%%%%
\section{Auger intensity}
\label{sec:auger}
The Auger process can be divided into two subprocesses. The first one is
the creation of a core hole with spin $\sigma_c$ at the lattice site
$i_c$ by absorbing an x-ray quantum for example. The second subprocess
is the radiationless decay of the core hole via ejecting an Auger
electron with spin $\sigma$ and momentum ${\bf k}$. Provided that the
life time of the core hole is large compared to typical relaxation times
of the valence electrons in the presence of the core hole, the two
subprocesses become independent from each other~\cite{AH83} (two-step
model). This implies the absence of any decay term in the Hamiltonian.
Consequently $n^c_{i_c \sigma_c}$ is a good quantum number,
$\left[H,n_{i_c\sigma_c}^c\right]_-=0$.  We can concentrate on the
second subprocess. Within the two-step model the initial state for the
Auger transition process is the ground state within the subspace
$\mathcal{H}^{\rm e}$ of the Hilbert space $\mathcal{H}$ that is built
up by all many-body states with $n^c_{i_c \sigma_c}=0$. To perform
thermodynamic averages in practice, one has to take into account this
restriction by introducing an additional Lagrange parameter.

The transition process itself is described by the transition
operator~\cite{PBBN93}
\begin{equation}
  T_{{\bf k}\sigma\sigma_c}=
  \sum_{L_1,L_2} M_{i_c{\bf k}}^{L_1L_2} 
  c_{i_cL_1\sigma}^\dag c_{i_cL_2\sigma_c}^\dag b_{i_c\sigma_c}
  \,.
  \label{eq:transop}
\end{equation}
${\bf k}$ and $\sigma$ denote the quantum-numbers of the Auger electron,
$\sigma_c$ the spin of the core state involved. The (intra-atomic)
Auger-matrix elements are given by
\begin{eqnarray}
\label{eq:augerme}
  M_{i_c{\bf k}}^{L_1L_2} 
  &=& 
  \langle i_c L_1, i_c L_2 
  |H_{\rm Coulomb}| {\bf k} , i_c \rangle \\
  &\propto &\int \!\!\! \int \! d^3 \! r_1 d^3 \!
  r_2 \,
  \overline{\Psi}_{L_1}({\bf r}_1-{\bf R}_{i_c})
  \overline{\Psi}_{L_2}({\bf r}_2-{\bf R}_{i_c})\nonumber\\
  &&\times\frac{1}{|{\bf r}_1-{\bf r}_2|}
  \Phi_{{\bf k}}({\bf r}_1)\phi({\bf r}_2-{\bf R}_{i_c})
  \nonumber
\end{eqnarray}
where $\Psi$ is the valence orbital, $\Phi$ the one-particle
wave-function of the Auger-electron and $\phi$ the core state. The bar
denotes complex conjugation.
\\
Following Ref.~\onlinecite{PBBN93} we consider the (retarded)
three-particle Green function, relevant for AES, which is defined as
(with the abbreviation ${\bf L}=(L_1,L_2)$ and ${\bf
  L}^\prime=(L^\prime_1,L^\prime_2)$)
\begin{eqnarray}
  \label{eq:3tgf}
  G_{{\bf k}\sigma\sigma_c}^{(3)}(E)
  &=&
  \langle\langle 
  T^\dag_{{\bf k}\sigma\sigma_c};T_{{\bf k}\sigma\sigma_c}
  \rangle\rangle_E
  \\ &=&
  \sum_{{\bf L},{\bf L}^\prime}
  \overline{M}_{i_c{\bf k}}^{\bf L}
  G^{(3),{\bf L}{\bf L}^\prime}_{i_c \sigma \sigma_c}(E)
  M_{i_c{\bf k}}^{{\bf L}^\prime}
  \nonumber
\end{eqnarray}
with
\begin{equation}
  G^{(3),{\bf L} {\bf L}^\prime}_{i_c \sigma \sigma_c}(E)=
  \langle\langle
  b_{i_c\sigma_c}^\dag c_{i_cL_2\sigma_c}c_{i_cL_1\sigma} ;
  c_{i_cL^\prime_1\sigma}^\dag c_{i_cL^\prime_2\sigma_c}^\dag b_{i_c\sigma_c}
  \rangle\rangle_{E}
  \,.
  \label{eq:sub3tgf}
\end{equation}
$\langle\langle\,.\,;\,.\,\rangle\rangle_E$ refers to Zubarev
Green functions~\cite{Zub60,Nol7}.  The AES intensity is then mainly
given by the three-particle spectral-density $A_{{\bf
    k}\sigma\sigma_c}^{(3)}(E) =-(1/\pi)\Im G_{{\bf
    k}\sigma\sigma_c}^{(3)}(E)$:
\begin{equation}
  I_{{\bf k}\sigma\sigma_c}(E+\epsilon_c-\mu)
  \propto \delta(E-E({\bf k})) A_{{\bf k}\sigma\sigma_c}^{(3)}(E)\,.
\end{equation}
Here $E({\bf k})$ is the dispersion of the Auger-electron.

In general the three-particle Green function will be a (complicated)
functional of one-particle Green functions. This functional represents
the direct correlations. In the following we concentrate on the indirect
correlations first, i.\,e.\ on the determination of the relevant
one-particle Green functions.
%%%%%%%%%%%%%%%%%%%%%%%%%%%%%%%%%%%%%%%%%%%%%%%%%%%%%%%%%%%%%%%%%%%%%%%
\section{Indirect Correlations}
\label{sec:indirect}
%%%%%%%%%%%%%%%%%%%%%%%%%%%%%%%%%%%%%
\subsection{Valence-band interaction}
\label{sec:vbi}
We consider the (retarded) one-particle valence-band Green-function
$\langle\langle c_{iL\sigma};c_{i^\prime L^\prime\sigma}^\dag
\rangle\rangle_E$.  Using a matrix notation with respect to the orbital
index $L=\{l,m\}$
\begin{equation}
 \left({\bf X}_{ii^\prime}\right)^{LL^\prime}
 =X_{ii^\prime}^{LL^\prime}
\end{equation}
and defining a lattice-Fourier transformation
\begin{equation}
  {\bf X}_{\bf k}=\frac{1}{N_{\rm s}}\sum_{i,i^\prime}
  e^{-i{\bf k}({\bf R}_i-{\bf R}_{i^\prime})}\,{\bf X}_{ii^\prime}
  \,,
\end{equation}
where $N_{\rm s}$ is the number of lattice sites, we get Dyson's
equation in the form
\begin{equation}
  \label{eq:vgf}
  {\bf G}_{ii^\prime \sigma}(E)=
  \frac{1}{N_{\rm s}}\sum_{\bf k}
  \frac{e^{i{\bf k}({\bf R}_i-{\bf R}_{i^\prime})}}
  {(E+\mu)\,{\bf S}_{\bf k}-
    {\bf t}_{\bf k}-{\bf \Sigma}_{{\bf k}\sigma}(E)}\,.
\end{equation}
Here the matrices ${\bf t}_{\bf k}$ and ${\bf S}_{\bf k}$ are the
Fourier-transformed hopping and overlap integrals of
Eqs.~(\ref{eq:hopping}) and (\ref{eq:overlap}), respectively.  ${\bf
  \Sigma}_{{\bf k}\sigma}(E)$ is the self-energy.  The one-particle
spectral density is given by
\begin{equation}
  {\bf A}_{ii^\prime \sigma}(E)=
  -\frac{1}{\pi}\Im {\bf G}_{ii^\prime \sigma}(E+i0^+)
  \,.
\end{equation}
The on-site terms of the spectral density are diagonal in the orbital
index as a consequence of lattice symmetries, and we have for the
orbital resolved QDOS
\begin{equation}
  \rho_{\sigma}^L(E)=A_{ii\sigma}^{LL}(E-\mu)
\end{equation}
where we dropped the site index. The total QDOS is obtained via
\begin{equation}
  \label{eq:tdos}
  \rho_{\sigma}(E)=\sum_{i^\prime}
  {\rm Tr} \left\{
    {\bf S}_{ii^\prime}{\bf A}_{{i^\prime i}\sigma}(E-\mu)\right\}
  \,.
\end{equation}
To calculate the self-energy, we use a standard approximation scheme,
the second order perturbation theory around the Hartree-Fock
solution~\cite{AGD,KM81,SAS92,SC91,PN97} (SOPT-HF).  It is known from
the single-band Hubbard model that the non-local terms of the SOPT-HF
self-energy rapidly decrease with increasing number of shells taken into
account~\cite{SC91,PN97}.  Furthermore, due to the band degeneracy there
is a much weaker ${\bf k}$-dependence of the SOPT-HF self-energy
compared to the single-band case~\cite{SAS92}. We may thus employ the
local approximation
\begin{equation}
  \Sigma_{ii^\prime \sigma}^{LL^\prime}(E)
  \approx\Sigma_{ii\sigma}^{LL^\prime}(E)\, \delta_{ii^\prime}
  =\Sigma_{\sigma}^L (E)
  \, \delta_{LL^\prime}\delta_{ii^\prime}
\end{equation}
As for the on-site Green function, lattice symmetries require the
on-site self-energy to be diagonal in the orbital index.

The Hartree-Fock (HF) contribution to the self-energy reads as
\begin{eqnarray}
  \label{eq:hfse}
  \Sigma_\sigma^{{\rm (HF),}L} &=&
  \sum_{L_1}
  \left\{ U_{LL_1LL_1}
    \left(n_{-\sigma}^{L_1}-n_{-\sigma}^{(0)L_1}\right)\right.+
    \\ & & \nonumber
    \left.\left(U_{LL_1LL_1}-U_{LL_1L_1L}\right)
    \left(n_{\sigma}^{L_1}-n_{\sigma}^{(0)L_1}\right)
  \right\}
\end{eqnarray}
where $n_{\sigma}^{L}=\langle n_{iL\sigma} \rangle$ denotes the
expectation value in the full model and $n_{\sigma}^{(0)L}$ the
expectation value of the band-structure calculation that stems from the
double counting correction in equation~(\ref{eq:hi}). Approximating the
self-energy by $\Sigma_\sigma^{{\rm (HF),}L}$, corresponds to the
LDA+$U$ approach~\cite{AAL97}. Here we additionally include the next
order in the interaction. The second-order contribution (SOC) to the
self-energy reads as
\begin{eqnarray}
  \lefteqn{\Sigma_\sigma^{{\rm (SOC)},L}(E)
    =\int\!\!\int\!\!\int \!\frac{dx\, dy\, dz}{E-x+y-z}}
  \\
  & &\times \left(f_-(x)f_-(-y)f_-(z)+f_-(-x)f_-(y)f_-(-z)\right)
  \nonumber\\
  & & \times\!\sum_{L_1,L_2,L_3}\!U_{LL_1L_3L_2}
  \,\widetilde{\rho}_{\sigma}^{L_3}(x)\,
  \Bigl\{U_{L_3L_2LL_1}\,
  \widetilde{\rho}_{-\sigma}^{L_1}(y)
  \widetilde{\rho}_{-\sigma}^{L_2}(z)  \nonumber \\
  & & \hspace{4em}
    +\left(U_{L_3L_2LL_1}-U_{L_3L_2L_1L}\right)\,
    \widetilde{\rho}_{\sigma}^{L_1}(y)
    \widetilde{\rho}_{\sigma}^{L_2}(z)\Bigr\}
  \nonumber
\end{eqnarray}
with the Fermi function $f_-(E)=(e^{\beta E}+1)^{-1}$ and where the
local HF spectral-density
$\widetilde{\rho}_{\sigma}^{L}(E)=A_{ii\sigma}^{{\rm (HF)},LL}(E)$ is
obtained by using the HF self-energy~(\ref{eq:hfse}) in
equation~(\ref{eq:vgf}).  The SOPT-HF self-energy
\begin{equation}
  {\bf \Sigma}_\sigma(E)=
  {\bf \Sigma}_\sigma^{\rm (HF)}+{\bf \Sigma}_\sigma^{\rm (SOC)}(E)
  \label{eq:vvse}
\end{equation}
determines the full Green function via equation~(\ref{eq:vgf}).
%%%%%%%%%%%%%%%%%%%%%%%%%%%%%%%%%%%%%%%%%
\subsection{Core-valence interaction}
%%%%%%%%%%%%%%%%%%%%%%%%%%%%%%%%%%%%%%%%%%%
\subsubsection{Valence-band Green function}
Let us now focus on the core hole screening in the initial state for
AES. The CV interaction and the presence of the core hole introduce an
additional (Hartree-like) term
\begin{equation}
  \Sigma_{i\sigma}^{{\rm (CV,e)},L}=-\delta_{ii_c} U^c_L
\end{equation}
to the valence-band self-energy which breaks the translational symmetry.
This term represents the core-hole potential at the lattice site $i_c$
where the core hole was created (the superscript ``e'' indicates
averaging in the restricted Hilbert-space $\mathcal{H}^{\rm e}$; see
section~\ref{sec:auger}). It is responsible for the screening.  The
valence-band self-energy in the presence of the core hole then reads
\begin{equation}
  {\bf \Sigma}_{i\sigma}^{\rm e}(E)=
  {\bf \Sigma}_{i\sigma}^{\rm (VV,e)}(E)+
  {\bf \Sigma}_{i\sigma}^{\rm (CV,e)}\,.
\end{equation}
${\bf \Sigma}_{\sigma}^{\rm (VV,e)}(E)$ incorporates the VV-correlation
effects and has the same structure as the self-energy~(\ref{eq:vvse})
for the translational invariant system. But in contrast to its
translational invariant counterpart, ${\bf \Sigma}_{\sigma}^{\rm
  (VV,e)}(E)$ is defined in terms of the Green functions in the presence
of the core hole.
\\
The valence band Green function ${\bf G}_{ii^\prime \sigma}^{\rm e}(E)$
in the presence of the core hole can be obtained by using Dyson's
equation in the form
\begin{eqnarray}
  \label{eq:vgfcd}
  \lefteqn{{\bf G}_{ii^\prime \sigma}^{\rm e}(E)
  ={\bf G}_{ii^\prime \sigma}(E)}\\
  &&+\sum_{j}{\bf G}_{ij\sigma}(E)
  \left({\bf \Sigma}_{j\sigma}^{\rm e}(E)-
    {\bf \Sigma}_{\sigma}(E)\right)
  {\bf G}_{ji^\prime \sigma}^{\rm e}(E)
  \,.\nonumber
\end{eqnarray}
In general ${\bf \Sigma}_{i\sigma}^{\rm e}(E)-{\bf
  \Sigma}_{\sigma}(E)\neq 0$ for a certain number of shells around the
core-hole site because the VV-correlation effects depend on the
occupation numbers, which as a consequence of the screening locally
differ from the translational invariant ones.  In the following we
assume complete screening, i.\,e.\ charge neutrality at the site $i_c$,
which is reasonable especially for 3d-transition metals because the
screening time scale is small compared to the life-time of the core
hole~\cite{SM98}. This implies ${\bf \Sigma}_{i\sigma}^{\rm e}(E)-{\bf
  \Sigma}_{\sigma}(E)$ to be small for all sites $i$ except for $i=i_c$.
Neglecting the terms for $i\neq i_c$ one can solve
equation~(\ref{eq:vgfcd})
\begin{eqnarray}
  \label{eq:gns0}
  \lefteqn{{\bf G}_{ii^\prime \sigma}^{\rm e}(E)
    ={\bf G}_{ii^\prime \sigma}(E)+{\bf G}_{ii_c\sigma}(E)}\\
  &&\times\left(\left({\bf \Sigma}_{i_c \sigma}^{\rm e}(E)
        -{\bf \Sigma}_{\sigma}(E)\right)^{-1}-{\bf G}_{i_ci_c\sigma}(E)
    \right)^{-1}{\bf G}_{i_ci^\prime \sigma}(E)
  \,.\nonumber
\end{eqnarray}
For $i=i^\prime=i_c$ one obtains the local screened Green-function at
the core-hole site:
\begin{equation}
  {\bf G}_{i_ci_c\sigma}^{\rm e}(E)=
  \frac{1}{{\bf G}^{-1}_{i_ci_c\sigma}(E)
    -\left({\bf \Sigma}_{i_c \sigma}^{\rm e}(E)
      -{\bf \Sigma}_{\sigma}(E)\right)}\,.
\end{equation}
The assumption of complete screening will be utilised as condition to
fix the CV-interaction parameter, which is taken to be the same for s-,
p- and d-orbitals ($U^c_L\equiv U_c$).
%%%%%%%%%%%%%%%%%%%%%%%%%%%%%%%%%%%%%%%%%%%%%%%%%%%%%%%%%%%%%%%%%%
\subsubsection{Core Green function}
To take into account the CV interaction for the core Green function
\begin{eqnarray}
  g_{i_c \sigma_c}(E)
  &=&\langle\langle b_{i_c \sigma_c}; b_{i_c \sigma_c}^\dag\rangle\rangle_E
  \\
  &=&\frac{1}{E+\mu -\epsilon_c -
    \Sigma_{i_c \sigma_c}^{c}(E)}
  \nonumber
\end{eqnarray}
one may calculate the core self-energy $\Sigma_{i_c \sigma_c}^{c}(E)$
using e.\,g.\ the SOPT-HF in the same way as for the valence-band Green
function. On the other hand it is believed that the core states are
influenced by other and presumably more important effects, such as
life-time effects~\cite{AH83}.  In fact, the core spectral density
obtained within SOPT-HF turns out to be dominated by a delta-peak that
is shifted by about 1\,eV below $\epsilon_c-\mu$. This does not affect
the Auger line shape.  Therefore, we assume for convenience the core
self-energy to be zero.  The spectral density becomes
\begin{equation}
  \label{eq:acore}
  a_{i_c \sigma_c}(E)
  =-\frac{1}{\pi}\Im g_{i_c \sigma_c}(E+i0^+)
  =\delta(E+\mu-\epsilon_c)
\,.
\end{equation}
%%%%%%%%%%%%%%%%%%%%%%%%%%%%%%%%%%%%%%%%%%%%%%%%%%%%%%%%%%%%%%%%%
\subsection{Results for Ni}
\begin{figure}
  \begin{center}
    \epsfig{file=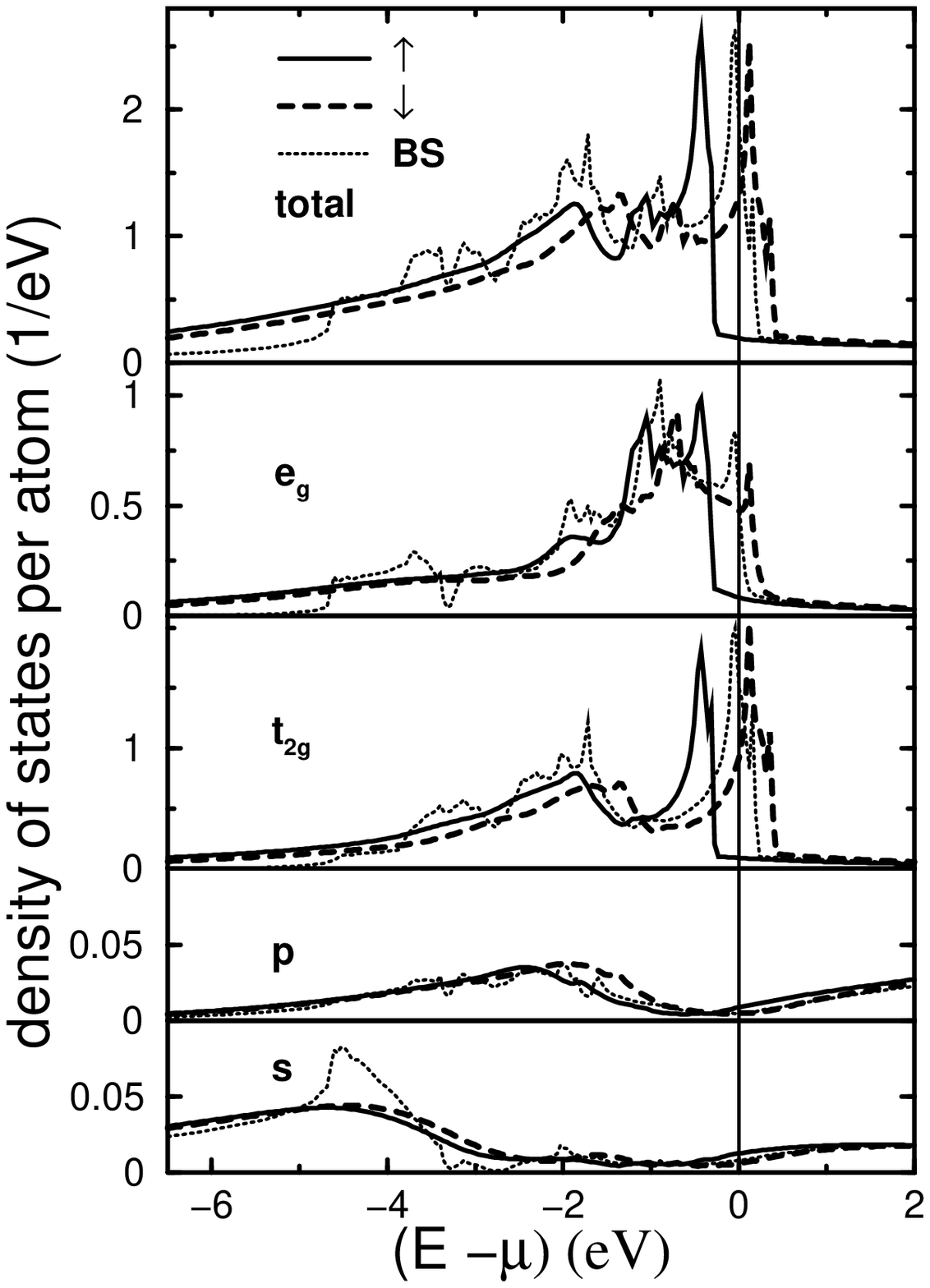, height=5.95cm}
    \epsfig{file=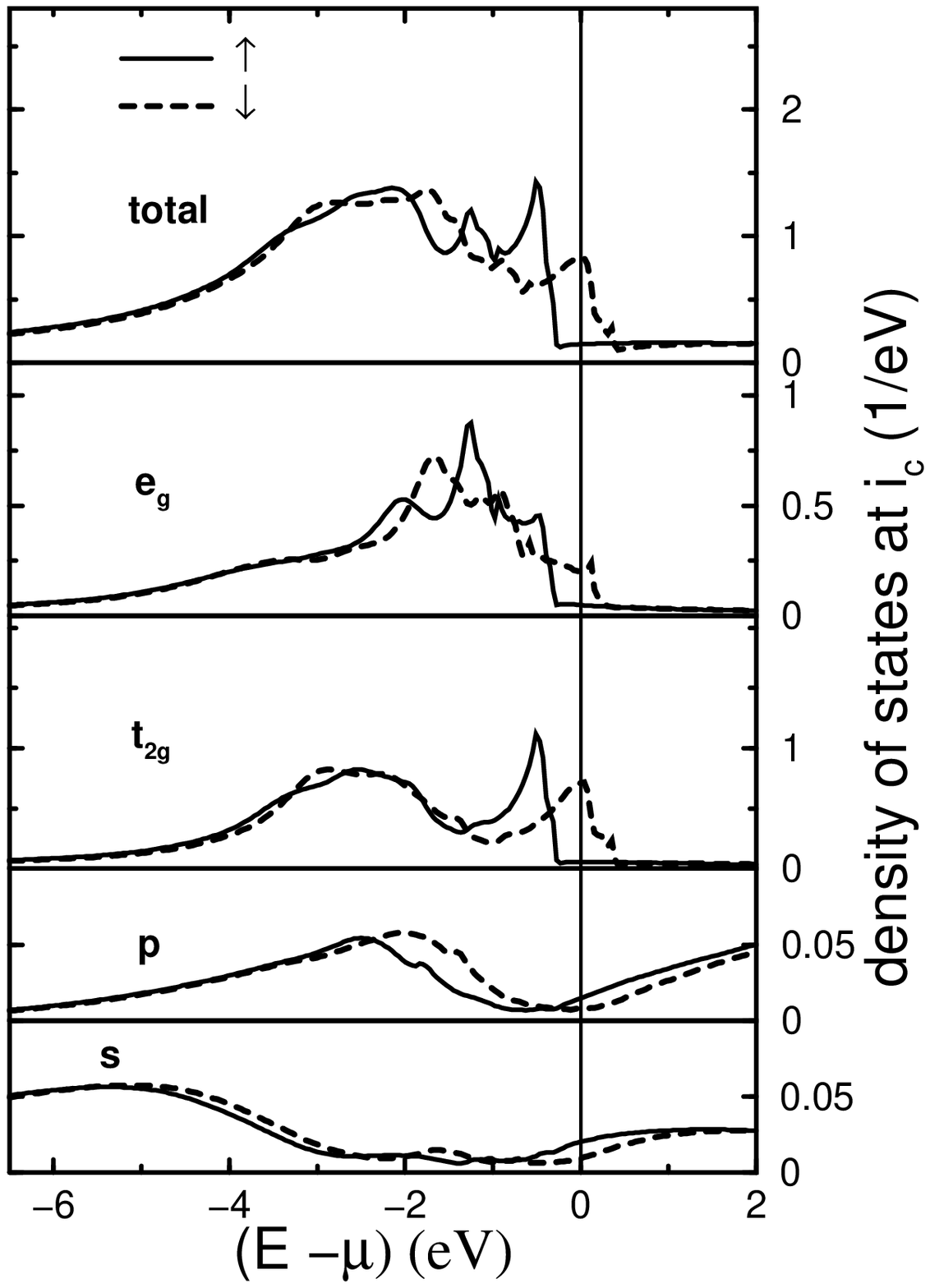, height=5.95cm}
    \caption{Spin-resolved QDOS per atom for s-, p-, ${\rm t_{2g}}$-,
      ${\rm e_g}$-states and total QDOS for $U=2.47$\,eV, $J=0.5$\,eV
      and $T=0$\,K.  Left: unscreened QDOS. Thin dotted line:
      tight-binding band-structure calculation~\cite{Pap86} for
      paramagnetic Ni.  Right: screened QDOS at the site $i_c$ in the
      presence of the core hole, $U_c=1.81$\,eV.}
    \label{fig:qdosju02}
  \end{center}
\end{figure}

\begin{figure}[b]
  \begin{center}
    \epsfig{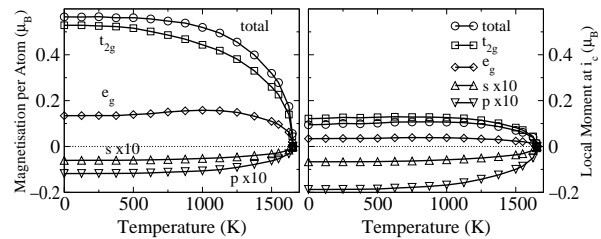}
    \caption{Left: magnetisation as function of temperature.
      Right: local moment as function of temperature in the presence of
      the core hole.}
    \label{fig:mcju02}
  \end{center}
\end{figure}

Before we discuss the results for fcc-Ni, we like to make a short remark
concerning the numerical evaluation of the theory.  The ${\bf k}$-sum
for the local Green function in equation~(\ref{eq:vgf}) was performed on
a mesh of 240 ${\bf k}$-points within the irreducible part of the
Brillouin zone using the tetrahedron method~\cite{LV84} generalised to
complex band structures, similar to that presented in
Ref.~\onlinecite{APKAK97}. The evaluation of the total
QDOS~(\ref{eq:tdos}) as well as the QDOS in the presence of the core
hole was done in ${\bf k}$-space. For the latter,
equation~(\ref{eq:vgfcd}) has to be used to perform the Fourier
transformation.
\\
The effective Slater integrals or, equivalently, the averaged direct and
exchange interaction parameters are chosen as $U=2.47$\,eV and
$J=0.5$\,eV.  This leads to a calculated magnetic moment per atom of
$m=0.56$\,$\mu_{\rm B}$ at $T=0$\,K which is the same as the measured
moment~\cite{HS73}.  With the ratio $J/U\approx 0.2$ we assume a typical
value for the late 3d transition metals.  The values given in the
literature, for instance $U=3.7$\,eV, $J=0.27$\,eV~\cite{SAS92} and
$U=2.97$\,eV, $J=0.8$\,eV~\cite{FOH97}, are of the same order of
magnitude but slightly overestimate the magnetic moment within the
present theory.
\\
The ``free'' DOS, used as starting point for our theory is shown on the
left of figure~\ref{fig:qdosju02} (thin dotted line) and corresponds to
tight binding band structure calculations~\cite{Pap86} for paramagnetic
fcc-Ni.
\\
The left-hand side of figure~\ref{fig:qdosju02} shows the QDOS per atom
for the model parameters given above.  As is known from the experiment,
Ni is a strong ferromagnet, i.\,e.\ the majority-spin states are fully
occupied.  The renormalisation effects of the a priori uncorrelated s-
and p-states seen in figure~\ref{fig:qdosju02} can be traced back to the
hybridisation with the d-states.
\\
Taking into account the presence of the core hole and following the
procedure to fix the CV interaction parameter as described above (charge
neutrality), leads to $U_c=1.81$\,eV. The corresponding ``screened''
QDOS is plotted on the right-hand side of figure~\ref{fig:qdosju02}.
The structure has changed remarkably. Spectral weight of the d-electrons
from the upper band edge is transferred to lower energies, the s- and
p-states are more populated, too. The screened magnetic moment at the
site $i_c$ ($m_{i_c}^{\rm e}=0.1\mu_{\rm B}$ at $T=0$\,K) is
considerably decreased since the local occupation is increased (see also
figure~\ref{fig:mcju02}).
\\
\begin{figure}[b]
  \begin{center}
    \epsfig{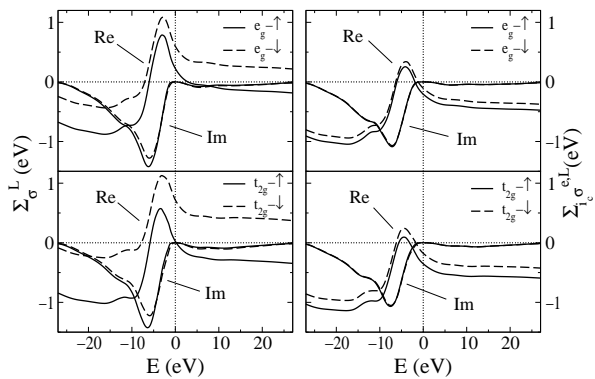}
    \caption{Real and imaginary part of the self-energy. Left:
      for the translational invariant system ($\Sigma_{\sigma}^L(E)$).
      Right: for the system in the presence of the core hole
      ($\Sigma_{i_c \sigma}^{{\rm e}L}(E)$)}
    \label{fig:secju02}
  \end{center}
\end{figure}
The left side of figure~\ref{fig:secju02} shows the self-energy
$\Sigma_{\sigma}^L(E)$. Within an energy range of about 1\,eV above and
below $E=0$\,eV one has $\Im \Sigma_\sigma^L(E) \propto E^2$. Thus we
have well-defined quasi-particles at the Fermi energy and their weight
\begin{equation}
  z^L_\sigma=
  \left|1-\frac{\partial \Re \Sigma^L_\sigma(E=0)}
      {\partial E}\right|^{-1}
\end{equation}
is 0.887 for the ${\rm t_{2g}}$-$\uparrow$-states and 0.893 for ${\rm
  t_{2g}}$-$\downarrow$-states.  For ${\rm e_g}$-states we find 0.878
and 0.883, respectively.  For energies above $-2$\,eV, where one finds
clearly distinguishable structures, a significant band-narrowing caused
by the real part of the self-energy is observed.  While the imaginary
part of the self-energy leads to strong damping effects in the QDOS
(figure~\ref{fig:qdosju02}, l.\,h.\,s.)  for energies below $-2$\,eV.
About $-6$\,eV below the Fermi energy where one expects the
``Ni-6-eV-satellite''~\cite{GBP77,SKO87,RM87} we find the largest
damping effects. However, we do not find the correlation-induced
6\,eV\,satellite. This is not surprising by applying the SOPT-HF or any
other finite-order perturbational approach~\cite{Cin79,Lie81,DJK98}.
For different interaction parameters (larger $U$, smaller $J$) a small
shoulder in the QDOS of the d-states is visible as was also mentioned in
Ref.~\onlinecite{KM81}.
\\
On the right-hand side of figure~\ref{fig:secju02} the self-energy in
the presence of the core hole $\Sigma_{i_c \sigma}^{{\rm e},L}(E)$ is
plotted.  Again there are well-defined quasi-particles, but with an
enhanced weight compared to the case where the core hole is absent
($z^{\rm t_{2g}}_{i_c \uparrow}=0.940$, $z^{\rm t_{2g}}_{i_c
  \downarrow}=0.934$, $z^{\rm e_{g}}_{i_c \uparrow}=0.937$, $z^{\rm
  e_{g}}_{i_c \downarrow}=0.936$).  The screened case behaves less
correlated than the unscreened case since here one is closer to the
limit of completely filled bands.
\\
Finally, we show the local magnetic moment per atom as a function of
temperature in figure~\ref{fig:mcju02}.  The magnetisation curves
(figure~\ref{fig:mcju02}, l.\,h.\,s.) have a Brillouin-function-like
form, except for the $e_g$-magnetisation which shows up a maximum at
$T\approx 1100$\,K.  This can be traced back to a transfer of charge
carriers from the ${\rm e_g}$-orbitals to the ${\rm t_{2g}}$-orbitals
with increasing temperature.  Because Ni is a strong ferromagnet the
charge-carrier transfer leads to an increase of the ${\rm
  e_g}$-magnetisation.  Contrary, the ${\rm t_{2g}}$-magnetisation is
decreased in addition to the usual temperature-induced depolarisation.
This leads to a temperature-dependent increase of the ratio $m_{\rm
  e_g}/(m_{\rm t_{2g}}+m_{\rm e_g})$ as it is known from polarised
neutron-scattering experiments~\cite{BDZ91,BDNZ92}. For $T=0$\,K this
ratio is $0.20$ and in good agreement with the measured value of
$0.19$~\cite{BDZ91,BDNZ92}.  As is observed experimentally, the s- and
p-states couple antiferromagnetically to the
d-states~\cite{BDZ91,BDNZ92}.  The Curie temperature turns out to be
$T_C=1655$\,K and is thereby about a factor 2.6 larger than the measured
value of 624\,K~\cite{LB19a}.  The large value for $T_C$ is probably due
to the mean-field character of the SOPT-HF. Note, however, that a simple
LDA+$U$ (HF) calculation for the same parameters $U$ and $J$ yields a
$T=0$ magnetisation 0.57\,$\mu_{\rm B}$ and a Curie temperature of
approximately 2500\,K.
\\
The local moment at the site $i_c$ as function of temperature is shown
on the right-hand side of figure~\ref{fig:mcju02}. Its d-contribution is
strongly reduced compared to the unscreened case while the s- and
p-moment is increased. The strong reduction of the total magnetic moment
will influence the spin polarisation of AES, since the ``screened'' QDOS
enters the Auger Green function~\ref{eq:3tgf}. Note that the total
magnetisation has to be calculated using equation~(\ref{eq:tdos}),
incorporating hybridisation with delocalised states.
%%%%%%%%%%%%%%%%%%%%%%%%%%%%%%%%%%%%%%%%%%%%%%%%%%%%%%%%%%%%%%%%%%%%%%
\section{Direct correlations}
\label{sec:direct}
To express the Auger intensity as a functional of the one-particle Green
functions we consider the diagrammatic expansion of the three-particle
Green function $G_{i_c \sigma \sigma_c}^{(3),{\bf L}{\bf L}^\prime}(E)$
(${\bf L}=(L_1,L_2)$).  Here we can restrict ourselves to the direct
diagrams and incorporate the exchange diagrams by introducing ``direct''
($D_{i_c{\bf k}}^{\bf L}=M_{i_c{\bf k}}^{L_1 L_2}$) and ``exchange''
Auger matrix elements ($E_{i_c{\bf k}}^{\bf L}=M_{i_c{\bf k}}^{L_2
  L_1}$).  The Auger intensity then reads as
\begin{eqnarray}
  \label{eq:augerint}
  \lefteqn{I_{{\bf k}\sigma\sigma_c}(E+\epsilon_c-\mu)\propto}\\&&
  \delta(E-E({\bf k}))\sum_{{\bf L},{\bf L}^\prime}
  \left(\overline{D}_{i_c{\bf k}}^{\bf L}
    -\delta_{\sigma\sigma_c}
    \overline{E}_{i_c{\bf k}}^{\bf L}\right)
  A_{i_c \sigma \sigma_c}^{(3),{\bf L}{\bf L}^\prime}(E)\,
  D_{i_c{\bf k}}^{{\bf L}^\prime}
  \,.\nonumber
\end{eqnarray}
The Auger matrix elements are taken to be constant. Following
Ref.~\onlinecite{KR97} we set:
\begin{equation}
  M_{i_c {\bf k}}^{L_1L_2}=\left\{%
    \begin{array}[c]{rl}
      1 & \mbox{for } L_1\leq L_2 \\
      -1 & \mbox{for } L_1> L_2\;.
    \end{array}\right.
  \label{eq:augerc}
\end{equation}
We thereby account for the singlet contributions, i.\,e.\ the holes in
the final state have opposite spin ($\sigma=-\sigma_c$), as well as for
the triplet contributions ($\sigma=\sigma_c$) to the Auger intensity.
The triplet contributions would be ignored if the Auger matrix elements
were chosen to be symmetric in the orbital index ($M_{i_c{\bf k}}^{L_1
  L_2}=M_{i_c{\bf k}}^{L_2 L_1}$) because the transition operator then
vanishes, as can be seen from equation~(\ref{eq:transop}).
%%%%%%%%%%%%%%%%%%%%%%%%%%%%%%%%%%%%
\subsection{VV ladder approximation}
\begin{figure}[b]
  \begin{center}
    \epsfig{file=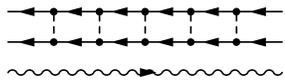, height=1cm}
    \caption{Typical diagram of the VV ladder. Solid line:
      renormalised valence-band propagator. Wiggly line:
      core-propagator. Dashed line: VV interaction.}
    \label{fig:2ladder}
  \end{center}
\end{figure}
We consider two different approximation schemes for the treatment of the
direct correlations. In the first approach, following
Refs.~\onlinecite{Cin79,TDDS81,DK84,Nol90,NGE91} we neglect the direct
CV correlations and treat the direct VV correlations by means of the VV
ladder approximation, which becomes exact in the limit of completely
filled or empty bands.  A typical diagram contributing to the VV ladder
is shown in figure~\ref{fig:2ladder}. The solid lines represent the
renormalised one-particle propagators of the valence-band while the
wiggly line is the one-particle core-propagator. The dashed line
corresponds to the VV interaction.  Summing up all diagrams of this
kind, yields the VV ladder approximation. The three-particle spectral
density is given by
\begin{eqnarray}
  \label{eq:3tsd}
  A_{i_c \sigma \sigma_c}^{(3),{\bf L}{\bf L}^\prime}(E)
  &=&
  \int \! dE^\prime \,
  A_{i_c \sigma \sigma_c}^{(2),{\bf L}{\bf L}^\prime}(E+E^\prime)
  \nonumber\\
  &&\times
  a_{i_c \sigma_c}(E^\prime)f_+(E+E^\prime)
  \\
  &\stackrel{(\ref{eq:acore})}{=}&
  A_{i_c \sigma \sigma_c}^{(2),{\bf L}{\bf L}^\prime}(E+\epsilon_c-\mu)
  f_+(E+\epsilon_c-\mu)
  \nonumber
\end{eqnarray}
where $f_+(E)=(e^{\beta E}-1)^{-1}$ is the Bose function.  The
two-particle valence-band spectral density $A_{i_c \sigma
  \sigma_c}^{(2),{\bf L}{\bf L}^\prime}(E)$ is obtained from the
corresponding two-particle Green function.  Using a matrix notation with
respect to ${\bf L}=\{L_1,L_2\}$ and ${\bf L}^\prime$, the two-particle
Green function reads as
\begin{equation}
  \label{eq:ladder}
  {\bf G}_{i_c \sigma \sigma_c}^{(2)}(E)
  ={\bf G}_{i_c \sigma \sigma_c}^{(2,0)}(E)
  \left(1-{\bf U}{\bf G}_{i_c \sigma \sigma_c}^{(2,0)}(E)\right)^{-1}
\end{equation}
with $U_{{\bf L}{\bf L}^\prime}=U_{L_1 L_2 L_1^\prime L_2^\prime}$.
$G_{i_c \sigma\sigma_c}^{(2,0),{\bf L}{\bf L}^\prime}(E)$ is the
two-particle Green function which has to be calculated form the
self-convolution of the partial QDOS
\begin{eqnarray}
  \label{eq:sc}
  A_{i_c \sigma \sigma_c}^{(2,0),{\bf L}{\bf L}^\prime}(E)
  &=&\delta_{{\bf L}{\bf L}^\prime}\int \! dE^\prime \, 
  \rho_{i_c\sigma}^{L_1}(E-E^\prime)
  \rho_{i_c\sigma_c}^{L_2}(E^\prime)
  \\&&\hspace{-1cm}\times
  (f_-(E^\prime-E)f_-(-E^\prime)-f_-(E-E^\prime)f_-(E^\prime))\nonumber
\end{eqnarray}
using the spectral representation:
\begin{equation}
  G_{i_c \sigma \sigma_c}^{(2,0),{\bf L}{\bf L}^\prime}(E)=
  \int \! dE^\prime \,
  \frac{A_{i_c \sigma \sigma_c}
    ^{(2,0),{\bf L}{\bf L}^\prime}(E^\prime)}{E-E^\prime}
  \,.
\end{equation}
In equation~(\ref{eq:ladder}) we have applied the local approximation,
i.\,e.\ only the on-site elements for $i=i_c$ are assumed to be non-zero.
This approximation is analogous to the local approximation for the
one-particle Green function in section~\ref{sec:indirect}.
\subsection{VV and CV correlations}
\begin{figure}[b]
  \begin{center}
    \epsfig{file=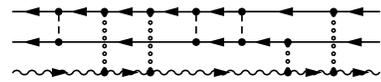, height=1cm}
    \caption{Typical diagram in the CVV ladder
      approximation. The notation is the same as in
      figure~\ref{fig:2ladder}. The additional CV interaction is
      represented by the dotted line.}
    \label{fig:3ladder}
  \end{center}
\end{figure}
\begin{figure}[t]
  \begin{center}
    \epsfig{file=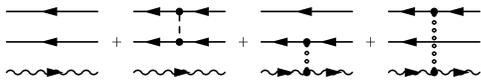, height=1cm}
    \caption{Diagrams up to first order in the VV and
      CV interaction (notation as in figure~\ref{fig:3ladder}).}
    \label{fig:firstorder}
  \end{center}
\end{figure}

A straightforward way to include the direct CV correlation on the same
level as the direct VV correlations has been discussed in
Refs.~\onlinecite{PBBN93,PBB94,PBNB94,PBBN95}. This leads to the
(``three-particle'') CVV ladder approximation. For the limiting case of
completely filled or empty bands the CVV ladder represents the exact
solution and recovers the VV ladder but shifted energetically by $2U_c$
due to the CV interaction~\cite{PBNB93}.  A typical diagram is shown in
figure~\ref{fig:3ladder} where the dotted line represents the CV
interaction.  The CVV ladder approximation leads to a coupled set of
Fredholm integral equations~\cite{PBNB93}. For a multi-band model,
however, the numerical evaluation is beyond our present computational
capacities. We therefore discuss a simpler approximation where only
diagrams up to first order in the direct correlations are retained (see
figure~\ref{fig:firstorder}).
%%%%%%%%%%%%%%%%%%%%%%%%%%%%%%%%%
\subsection{Results for Ni}
\begin{figure}[htbp]
  \begin{center}
    \epsfig{file=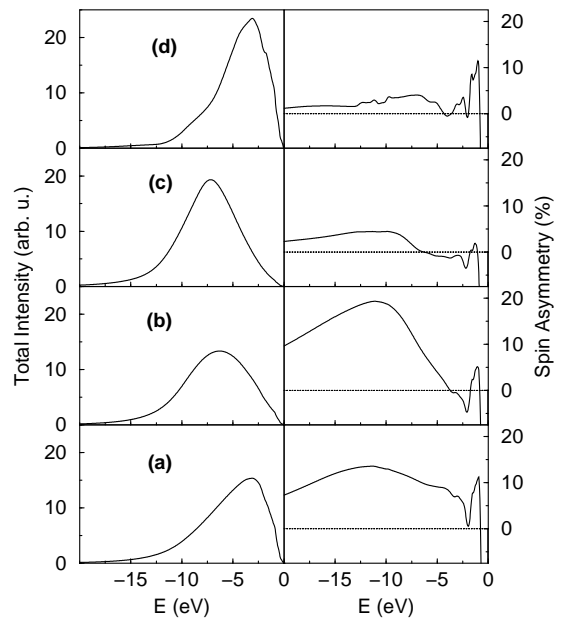,
      width=0.8\linewidth}
    \caption{Left: total intensities $I_{\uparrow}(E)+I_{\downarrow}(E)$.
      Right: spin-asymmetry $(I_{\uparrow}(E)-I_{\downarrow}(E))/
      (I_{\uparrow}(E)+I_{\downarrow}(E))$.  (a) self-convolution
      without screening of the core hole in the initial state; (b) VV
      ladder without core hole screening; (c) screened VV ladder; (d)
      direct VV and CV correlations included up to first order
      (figure~\ref{fig:firstorder}).}
    \label{fig:aest0}
  \end{center}
\end{figure}

The calculated Auger spectra for Ni resulting from different
approximations are shown in figure~\ref{fig:aest0}. The core hole is
assumed to be unpolarised (non-resonant~\cite{AMTL87} process).  The
intensities for core spin $\sigma_c$ and $-\sigma_c$ are added
incoherently:
$I_{\sigma}(E)=(I_{\sigma\sigma_c}(E)+I_{\sigma-\sigma_c}(E))/2$.
However, the Auger intensity is still spin-dependent due to the
ferromagnetic order in Ni.
\\
In figure~\ref{fig:aest0} we plotted the total Auger-intensity
$I_{\uparrow}(E)+I_{\downarrow}(E)$ on the left and the spin asymmetry
$(I_{\uparrow}(E)-I_{\downarrow}(E))/(I_{\uparrow}(E)+I_{\downarrow}(E))$
on the right.  Part (a) shows the result of the self-convolution model
(equations (\ref{eq:3tsd}) and (\ref{eq:sc}) inserted in
(\ref{eq:augerint})), i.\,e.\ the self-convolution of the occupied QDOS
in figure~\ref{fig:qdosju02} (l.\,h.\,s.). Direct correlations and
core-hole screening are neglected altogether. Part (b) corresponds to
the VV ladder approximation starting from the (unscreened) QDOS. Taking
additionally into account the screening effects introduced by the
presence of the core hole in the initial state results in (c).  The
spectrum obtained by the first order in the direct VV and CV
correlations (see figure~\ref{fig:firstorder}) and by the screened QDOS
is plotted in part (d).
\\
As one can see in the plots on the left-hand side, the VV interaction is
too weak to produce bound states, no sharp satellite appears, and the
spectra appear to be band-like. Compared to the self-convolution (a),
however, a considerable shift to lower energies is observed in (b). This
shift results from the direct correlations between the two final-state
holes in the valence band.  In (c) the main peak is shifted to still
lower energies. This is an effect of the core-hole screening in the
initial state and can be traced back to the redistribution of spectral
weight in the screened QDOS (figure~\ref{fig:qdosju02}). Compared to (a)
and (b), the total AES intensity is clearly increased in (c). Again this
is a consequence of the core-hole screening since the number of occupied
states available for the Auger process is increased
(figure~\ref{fig:qdosju02}, r.\,h.\,s.). The spectrum shown in (d) not
only includes the initial-state core-hole screening but the final-state
effects due to the destruction of the core hole. Compared to (c), where
the initial state is described in the same way, these effects result in
a strong shift of the main peak to higher energies. This shift almost
exactly compensates the shifts to lower energies that are due to direct
VV correlations (b) and the core-hole screening (c). However, a weak
shoulder at about $E=-8.5$\,eV remains in the spectrum (d).
\\
In all cases there is a high spin asymmetry (up to $-50$\,\%) for
energies between approximately $-0.8$\,eV and 0. This is a consequence
of the fact that Ni is a strong ferromagnet. There are almost no
$\uparrow$-electrons above about $-0.4$\,eV (see
figure~\ref{fig:qdosju02}) that can participate in the Auger-process.
The main contribution to the intensity is therefore due to triplet
configurations where the two final-state holes or, equivalently,
core-hole and Auger electron have spin\,$\downarrow$.  However, the
intensity is very small in this energy region. By taking into account
the screening of the core hole in the initial state (compare (b) and
(c)) the spin asymmetry is reduced over the whole energy range which
essentially is the same effect as the reduction of the local magnetic
moment at $i_c$ caused by the presence of the core hole
(figure~\ref{fig:mcju02}). The total spin-polarisation
\begin{equation}
  P=\frac{\int\!dE\,I_{\uparrow}(E) - \int\!dE\,I_{\downarrow}(E)}
  {\int\!dE\,I_{\uparrow}(E) + \int\!dE\,I_{\downarrow}(E)}
\end{equation}
in the case (d) is 2.6\,\% and 1.6\,\% for (c). Both values are close to
the experimental value~\cite{AMTL87} of 2\,\% for the
$M_1M_{45}M_{45}$-process. The cases (a) and (b) with a polarisation of
8.7\,\% and 9.3\,\%, respectively, overestimate the total
spin-polarisation compared with the experimental value.
\\
\begin{figure}[t]
  \begin{center}
    \epsfig{file=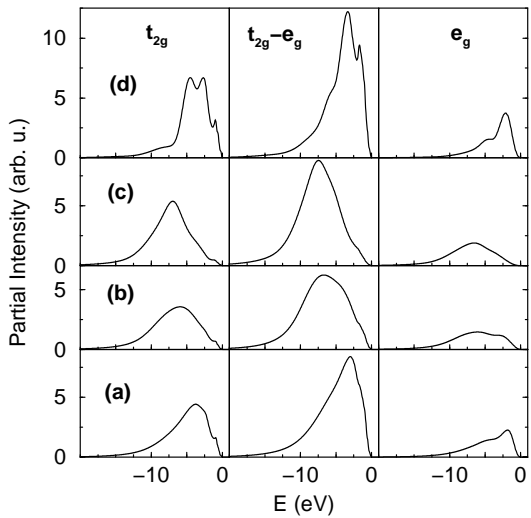,
      width=0.8\linewidth}
    \caption{Contributions to the total Auger intensity
      (figure~\ref{fig:aest0}) of processes involving ${\rm
        t_{2g}}$-electrons only (left), ${\rm e_g}$-electrons only
      (right) and both kinds of d-electrons (middle).}
    \label{fig:aespt0}
  \end{center}
\end{figure}
For the calculation of the orbitally resolved contributions to the Auger
intensity we may restrict the summation in equation (\ref{eq:augerint})
to orbital indices ($L_1,L_2,L_1^{\prime},L_2^{\prime}$) belonging to
${\rm t_{2g}}$(${\rm e_g}$)-character only. The resulting contributions
are shown on the left (right) of figure~\ref{fig:aespt0}.  The
contributions due to the remaining terms are plotted in the middle.
\\
In all cases (a)--(b) the ${\rm t_{2g}}$ contributions are clearly
stronger compared with the ${\rm e_g}$ contributions. The ratio between
the ${\rm t_{2g}}$ and ${\rm e_g}$ partial intensities corresponds to
the different degeneracies. Comparing the cases (a)--(d), we notice
that there are essentially the same trends in the partial intensities as
for the total intensities, and the discussion is the same as above. The
line shape in case (d), however, shows up some fine structure,
especially in the ${\rm t_{2g}}$ partial intensity, which is not that
pronounced in the total intensity. The shoulder at the low-energy tail
of (d) is due to direct VV correlations and may be interpreted as a hint
for the formation of a bound state of the final holes. Except for this
shoulder, a surprising similarity between (d) and (a) is noticed, even
for the orbital resolved spectra. This might also be due to the small
number of diagrams taken into account. However, the cancellation of
effects according to different interactions (VV and CV) was also pointed
out in Ref.~\onlinecite{SM98}.
%%%%%%%%%%%%%%%%%%%%%%%%%%
\section{Summary}
\label{sec:sum}
In this paper we have investigated electron-correlation effects on the
Auger line shape of Ni as an example of a ferromagnetic 3d-transition
metal.  The starting point is a realistic set of hopping and overlap
parameters taken from tight-binding band-structure calculations.  We
additionally consider a strongly screened on-site Coulomb interaction
between the rather localised 3d electrons. The respective Coulomb-matrix
elements are expressed in terms of effective Slater integrals.  Choosing
a non-orthogonal basis, a distinction between the different
angular-momentum characters of the valence orbitals is possible. This is
necessary for the precise definition of the Coulomb-interaction part in
the (multi-band Hubbard) Hamiltonian and also facilitates the
interpretation of the Auger spectra.  Furthermore, we account for the
core-valence interaction which is responsible for the screening of the
core hole in the initial state of AES and for the sudden response of the
valence electrons due to the destruction of the core hole in the final
state.

Within a diagrammatic approach, the indirect and the direct correlations
can be studied separately.  The indirect correlations have been treated
by second-order perturbation theory around the Hartree-Fock solution
(SOPT-HF).  The VV interaction parameters are fixed by assuming a ratio
$J/U\approx 0.2$ and by fitting the experimentally observed magnetic
moment for $T=0$\,K, leading to $U=2.47$\,eV and $J=0.5$\,eV. This is
equivalent to an intra-orbital interaction of $U_{LLLL}=3.07$\,eV
($L=\{2,m_2\}$). The resulting Curie temperature within the theory
presented here is by a factor 2.6 larger than the experimentally
observed one but considerably lower than the LDA+$U$ (Hartree-Fock)
value.
\\
The core-valence interaction leads to a breakdown of the translational
invariance in the initial state for AES. The interaction parameter
$U_c=1.81$\,eV is fixed be requiring charge neutrality at the site $i_c$
where the core hole is created.  The screening of the additional
core-hole potential causes a transfer of spectral weight below the Fermi
energy and thus a considerable reduction of the local magnetic moment.
However, the local magnetic moment is finite since the s- and
p-electrons also contribute to the screening.

To study the direct correlations we have used two different approaches.
The first is the VV ladder approximation, which results in a band-like
Auger spectrum with a single maximum.  In a second approach we have
summed up the first-order diagrams with respect to the VV and CV
interaction. The resulting spectrum shows up a shoulder at the lower
tail due to the VV interaction.  Otherwise the line shape is very
similar to that obtained by the self-convolution of the unscreened QDOS.
\\
As far as concerns the line shape, we conclude that the different
correlation effects, VV and CV correlations in the initial and the final
state, nearly cancel. However, a strong effect of electron correlations
has been found in the orbitally resolved partial intensities and
particularly for the spin-asymmetry.

The calculated spin polarisation is in a good agreement with the
measured one of the $M_1M_{45}M_{45}$-process~\cite{AMTL87}. This
corresponds to the excitation of a not too deep lying core level,
i.\,e.\ the two-step model should be applicable. The reduced but finite
spin polarisation of the $M_1M_{45}M_{45}$-process (as compared to the
band polarisation) can be explained by effects of core-hole screening
rather than by a core-hole polarisation, caused by a resonant excitation
of the core electron into the valence band~\cite{AMTL87}.  Future work
will show whether these findings also apply to other 3d-transition
metals.
%%%%%%%%%%%%%%%%%%%%%%%%%%%%%%%%%%%%%%%%%%%%%%%%%
\acknowledgements Financial support of the \textit{Deutsche
  Forschungsgemeinschaft} within the project No.~158/5-1 is greatfully
acknowledged.  The numerical calculations were performed on the CrayT3E
at the \textit{Konrad-Zuse-Zentrum f\"ur Informationstechnologie Berlin
  (ZIB)}.
%%%%%%%%%%%%%%%%%%%%%%%%%%%%%%%%%%%%%%%%%%%%%%%%
\appendix
%%%%%%%%%%%%%%%%%%%%%%%%%%%%%%%%%%%%%%%%%%%%%%%%%%%%%%
% For unique references the following macro is needed:
\renewcommand{\theequation}{\Alph{section}.\arabic{equation}}
%%%%%%%%%%%%%%%%%%%%%%%%%%%%%%%%%%%%%%%%%%%%%%%%%%%%%%
\section{Non-Orthogonal Basis-Set}
\label{sec:nobs}
There are several advantages for using the non-orthogonal basis set
$\{|iL\sigma\rangle\}$.  The Slater-Koster parameters~\cite{SK54} for
the two-center approximation (used here) are much more accurate for a
non-orthogonal basis set compared to an orthonormal one~\cite{Pap86}.
Secondly, the non-orthogonal basis (LCAO basis) is built up from
quasi-atomic orbitals. One therefore knows the behaviour of the basis
states under symmetry operations belonging to the $O_h$-group which
eventually results in the fact that local quantities, e.\,g.\ the on-site
Green function, are diagonal in the orbital index.  Furthermore, the
Coulomb matrix elements can be calculated in a highly symmetric way by
using $3j$-symbols in combination with a transformation from spheric to
cubic harmonics. The unknown radial parts of the basis are parametrised
by the effective Slater integrals ($F^0$, $F^2$, $F^4$).  Finally, one
can ensure that the Coulomb interaction acts between 3d-electrons only.

For the formalism of second quantisation, for the many-body and
Green-function theory, for the proof of Wick's theorem and thus the
development of the diagram technique, however, the use of an orthonormal
set of one-particle basis states is necessary. Therefore, it is convenient to derive
all expressions that refer to the non-orthogonal basis,
\begin{equation}
  \langle \alpha|\beta\rangle
  =S_{\alpha\beta}\,,
\end{equation}
by a (non-unitary) L\"owdin transformation~\cite{Loe50} from a related
set of orthogonal one-particle basis states:
\begin{equation}
  \label{eq:ob}
  |\widetilde{\alpha}\rangle\stackrel{\rm Def.}{=}\sum_\beta |\beta\rangle
  S_{\beta\alpha}^{-1/2}.
\end{equation}
The overlap matrix ${\bf S}$ is Hermitian,
$\{|\widetilde{\alpha}\rangle\}$ indeed represents an orthonormal and
complete basis set.  The completeness relation can be written as
\begin{eqnarray}
  \label{eq:complete}
  1&=&\sum_{\alpha}|\widetilde{\alpha}\rangle\langle\widetilde{\alpha}|
  =\sum_{\alpha\beta\gamma} |\beta\rangle\,
  S_{\beta\alpha}^{-1/2}S_{\alpha\gamma}^{-1/2}\,\langle\gamma|
  \nonumber\\
  &=&\sum_{\alpha\beta}|\alpha\rangle\,S_{\alpha\beta}^{-1}\,\langle\beta|\,.
\end{eqnarray}
Annihilation (and creation) operators referring to the non-orthogonal basis may be
defined as:
\begin{equation}
\label{eq:oc}
  c_\alpha\stackrel{\rm Def.}{=}\sum_\beta
  S_{\alpha\beta}^{-1/2}\,\widetilde{c}_\beta\,.
\end{equation}
It is instructive to see how the creation operator acts on the
vacuum state:
\begin{eqnarray}
  c_{\alpha}^{\dag} |0\rangle &=&
  \sum_{\beta} \widetilde{c}_\beta^{\dag} S_{\beta\alpha}^{-1/2}
  |0\rangle
  =\sum_\beta |\widetilde{\beta}\rangle S_{\beta\alpha}^{-1/2}
  \nonumber\\
  &=&\sum_{\beta\gamma} |\gamma\rangle
  S_{\gamma\beta}^{-1/2}S_{\beta\alpha}^{-1/2}
  =\sum_\beta |\beta\rangle S_{\beta\alpha}^{-1}
  \,.
  \label{eq:constr}
\end{eqnarray}
Furthermore, one gets from the transformations (\ref{eq:ob}) and
(\ref{eq:oc}):
\begin{equation}
  \label{eq:nosum}
  \sum_\alpha |\widetilde{\alpha}\rangle \, \widetilde{c}_\alpha
  =\sum_{\alpha\beta\gamma} \, |\beta\rangle
  S_{\beta\alpha}^{-1/2}S_{\alpha\gamma}^{1/2}
  \, c_\gamma
  =\sum_\alpha |\alpha\rangle \, c_\alpha\,.
\end{equation}
Thus, an operator in second quantisation has the same structure for
both, the orthonormal and the non-orthogonal basis set. A one-particle
operator $O$, for example, reads as:
\begin{eqnarray}
  O&=&
  \sum_{\alpha,\alpha^\prime}
  \widetilde{c}_\alpha^\dag\,
  \langle\widetilde{\alpha}|o|\widetilde{\alpha}^\prime\rangle
  \,\widetilde{c}_{\alpha^\prime}
  \nonumber\\&\stackrel{\rm (\ref{eq:nosum})}{=}&
  \sum_{\alpha,\alpha^\prime}
  c_\alpha^\dag\,
  \langle\alpha|o|\alpha^\prime\rangle
  \,c_{\alpha^\prime}\,.
\end{eqnarray}
The non-orthogonal Green functions are defined as in the orthonormal
case, e.\,g.\  $G_{\alpha\alpha^\prime}(E)=\langle\langle
c_{\alpha};c_{\alpha^\prime}^\dag \rangle\rangle$ for the one-particle
Green function. For example, using the non-orthogonal version of the
fundamental anti-commutation rules
($[c_\alpha,c_{\alpha^\prime}^\dag]_+=S_{\alpha\alpha^\prime}^{-1}$, see
also Eq.~(\ref{eq:acr})) and the equation of motion, the non-interacting
Green function turns out to be:
\begin{equation}
  {\bf G}^0(E)=\left((E+\mu)\,{\bf S}-{\bf T}\right)^{-1}
\end{equation}
where we used the matrix notation $({\bf
  G})_{\alpha\alpha^\prime}=G_{\alpha\alpha^\prime}$ etc. For the
interacting Green function one has (compare with Eq.~(\ref{eq:vgf})):
\begin{equation}
  {\bf G}(E)=\left((E+\mu)\,{\bf S}-{\bf T}-{\bf \Sigma}(E)\right)^{-1}
  \,.
\end{equation}

In the same way as for the examples given, one may use Wick's theorem,
develop the diagram technique etc.
%%%%%%%%%%%%%%%%%%%%%%%%%%%%


\begin{thebibliography}{10}
\providecommand*{\bibinfo}[2]{#2}
\providecommand*{\eprint}[1]{#1}
\providecommand*{\url}[1]{#1}
\bibitem{Fug81}
\bibinfo{author}{J.~C. Fuggle}, \bibinfo{title}{\emph{Electron Spectroscopy:
  Theory, Techniques and Applications}} (\bibinfo{publisher}{Academic}, London,
  \bibinfo{year}{1981}), \bibinfo{volume}{vol.~4}, \bibinfo{pages}{p.~85}.
\bibitem{WM81}
\bibinfo{author}{R.~Weissmann} and \bibinfo{author}{K.~M{\"u}ller},
  \bibinfo{journal}{Surf. Sci. Rep.} \bibinfo{volume}{\textbf{105}},
  \bibinfo{pages}{251} (\bibinfo{date}{1981}).
\bibitem{AH83}
\bibinfo{author}{C.-O. Almbladh} and \bibinfo{author}{L.~Hedin},
  \bibinfo{title}{\emph{Handbook on Synchrotron Radiation}}
  (\bibinfo{publisher}{North-Holland}, Amsterdam, \bibinfo{year}{1983}),
  \bibinfo{volume}{vol.~1b}, \bibinfo{pages}{p. 607}.
\bibitem{Wei84}
\bibinfo{author}{P.~Weightman}, \bibinfo{title}{\emph{Electronic Properties of
  surfaces}} (\bibinfo{publisher}{Adam Hilger}, Bristol, \bibinfo{year}{1984}),
  \bibinfo{pages}{p. 135}.
\bibitem{WAC88}
\bibinfo{author}{S.~B. Whitefield}, \bibinfo{author}{G.~B. Armen},
  \bibinfo{author}{R.~Carr}, \bibinfo{author}{J.~C. Levin}, and
  \bibinfo{author}{B.~Crasemann}, \bibinfo{journal}{Phys. Rev. A}
  \bibinfo{volume}{\textbf{37}}(2), \bibinfo{pages}{419}
  (\bibinfo{date}{1988}).
\bibitem{SCSG89}
\bibinfo{author}{D.~D. Sarma}, \bibinfo{author}{C.~Carbone},
  \bibinfo{author}{P.~Sen}, and \bibinfo{author}{W.~Gudat},
  \bibinfo{journal}{Phys. Rev. B} \bibinfo{volume}{\textbf{40}}(18),
  \bibinfo{pages}{12542} (\bibinfo{date}{1989}).
\bibitem{Ram91}
\bibinfo{author}{D.~E. Ramaker}, \bibinfo{journal}{Crit. Rev. Solid State
  Mater.} \bibinfo{volume}{\textbf{17}}, \bibinfo{pages}{211}
  (\bibinfo{date}{1991}).
\bibitem{SRC93}
\bibinfo{author}{D.~D. Sarma}, \bibinfo{author}{S.~R. Barman},
  \bibinfo{author}{R.~Cimino}, \bibinfo{author}{C.~Carbone},
  \bibinfo{author}{P.~Sen}, \bibinfo{author}{A.~Roy},
  \bibinfo{author}{A.~Chainani}, and \bibinfo{author}{W.~Gudat},
  \bibinfo{journal}{Phys. Rev. B} \bibinfo{volume}{\textbf{48}}(10),
  \bibinfo{pages}{6822} (\bibinfo{date}{1993}).
\bibitem{Lan53}
\bibinfo{author}{J.~J. Lander}, \bibinfo{journal}{Phys. Rev.}
  \bibinfo{volume}{\textbf{91}}, \bibinfo{pages}{1382} (\bibinfo{date}{1953}).
\bibitem{Pow73}
\bibinfo{author}{C.~J. Powell}, \bibinfo{journal}{Phys. Rev. Lett.}
  \bibinfo{volume}{\textbf{30}}, \bibinfo{pages}{1179} (\bibinfo{date}{1973}).
\bibitem{HWMR88}
\bibinfo{author}{G.~H{\"o}rmandinger}, \bibinfo{author}{P.~Weinberger},
  \bibinfo{author}{P.~Marksteiner}, and \bibinfo{author}{J.~Redinger},
  \bibinfo{journal}{Phys. Rev. B} \bibinfo{volume}{\textbf{38}}(2),
  \bibinfo{pages}{1040} (\bibinfo{date}{1988}).
\bibitem{KR97}
\bibinfo{author}{Y.~Kucherenko} and \bibinfo{author}{P.~Rennert},
  \bibinfo{journal}{J. Phys.: Condens. Matter} \bibinfo{volume}{\textbf{9}},
  \bibinfo{pages}{5003} (\bibinfo{date}{1997}).
\bibitem{Cin77}
\bibinfo{author}{M.~Cini}, \bibinfo{journal}{Solid State Commun.}
  \bibinfo{volume}{\textbf{24}}, \bibinfo{pages}{681} (\bibinfo{date}{1977}).
\bibitem{Saw77}
\bibinfo{author}{G.~A. Sawatzky}, \bibinfo{journal}{Phys. Rev. Lett.}
  \bibinfo{volume}{\textbf{39}}(8), \bibinfo{pages}{504}
  (\bibinfo{date}{1977}).
\bibitem{SL80}
\bibinfo{author}{G.~A. Sawatzky} and \bibinfo{author}{A.~Lenselink},
  \bibinfo{journal}{Phys. Rev. B} \bibinfo{volume}{\textbf{21}}(5),
  \bibinfo{pages}{1790} (\bibinfo{date}{1980}).
\bibitem{Cin78}
\bibinfo{author}{M.~Cini}, \bibinfo{journal}{Phys. Rev. B}
  \bibinfo{volume}{\textbf{17}}(6), \bibinfo{pages}{2788}
  (\bibinfo{date}{1978}).
\bibitem{NGE92}
\bibinfo{author}{W.~Nolting}, \bibinfo{author}{G.~Geipel}, and
  \bibinfo{author}{K.~Ertl}, \bibinfo{journal}{Phys. Rev. B}
  \bibinfo{volume}{\textbf{45}}(11), \bibinfo{pages}{5790}
  (\bibinfo{date}{1992}).
\bibitem{PBNB93}
\bibinfo{author}{M.~Potthoff}, \bibinfo{author}{J.~Braun},
  \bibinfo{author}{W.~Nolting}, and \bibinfo{author}{G.~Borstel},
  \bibinfo{journal}{J. Phys.: Condens. Matter} \bibinfo{volume}{\textbf{5}},
  \bibinfo{pages}{6879} (\bibinfo{date}{1993}).
\bibitem{NBDF89}
\bibinfo{author}{W.~Nolting}, \bibinfo{author}{W.~Borgie{\l}},
  \bibinfo{author}{V.~Dose}, and \bibinfo{author}{T.~Fauster},
  \bibinfo{journal}{Phys. Rev. B} \bibinfo{volume}{\textbf{40}}(7),
  \bibinfo{pages}{5015} (\bibinfo{date}{1989}).
\bibitem{NVF95}
\bibinfo{author}{W.~Nolting}, \bibinfo{author}{A.~Vega}, and
  \bibinfo{author}{T.~Fauster}, \bibinfo{journal}{Z. Phys. B}
  \bibinfo{volume}{\textbf{96}}, \bibinfo{pages}{357} (\bibinfo{date}{1995}).
\bibitem{PWN97}
\bibinfo{author}{M.~Potthoff}, \bibinfo{author}{T.~Wegner}, and
  \bibinfo{author}{W.~Nolting}, \bibinfo{journal}{Phys. Rev. B}
  \bibinfo{volume}{\textbf{55}}, \bibinfo{pages}{16132} (\bibinfo{date}{1997}).
\bibitem{KK97}
\bibinfo{author}{H.~Kajueter} and \bibinfo{author}{G.~Kotliar},
  \bibinfo{journal}{Int. J. Mod. Phys. B} \bibinfo{volume}{\textbf{11}},
  \bibinfo{pages}{729} (\bibinfo{date}{1997}).
\bibitem{LK98}
\bibinfo{author}{A.~I. Lichtenstein} and \bibinfo{author}{M.~I. Katsnelson},
  \bibinfo{journal}{Phys. Rev. B} \bibinfo{volume}{\textbf{57}}(12),
  \bibinfo{pages}{6884} (\bibinfo{date}{1998}).
\bibitem{HV98}
\bibinfo{author}{K.~Held} and \bibinfo{author}{D.~Vollhardt},
  \bibinfo{journal}{Eur. Phys. J. B} \bibinfo{volume}{\textbf{5}}(3),
  \bibinfo{pages}{473} (\bibinfo{date}{1998}).
\bibitem{MV89}
\bibinfo{author}{W.~Metzner} and \bibinfo{author}{D.~Vollhardt},
  \bibinfo{journal}{Phys. Rev. Lett.} \bibinfo{volume}{\textbf{62}},
  \bibinfo{pages}{324} (\bibinfo{date}{1989}).
\bibitem{KL99}
\bibinfo{author}{M.~I. Katsnelson} and \bibinfo{author}{A.~I. Lichtenstein},
  \bibinfo{journal}{J. Phys.: Condens. Matter} \bibinfo{volume}{\textbf{11}},
  \bibinfo{pages}{1037} (\bibinfo{date}{1999}).
\bibitem{SM98}
\bibinfo{author}{D.~D. Sarma} and \bibinfo{author}{P.~Mahadevan},
  \bibinfo{journal}{Phys. Rev. Lett.} \bibinfo{volume}{\textbf{81}}(8),
  \bibinfo{pages}{1658} (\bibinfo{date}{1998}).
\bibitem{Drc89}
\bibinfo{author}{V.~Drchal}, \bibinfo{journal}{J. Phys.: Condens. Matter}
  \bibinfo{volume}{\textbf{1}}, \bibinfo{pages}{4773} (\bibinfo{date}{1989}).
\bibitem{KD92}
\bibinfo{author}{M.~Kotrla} and \bibinfo{author}{V.~Drchal},
  \bibinfo{journal}{J. Phys.: Condens. Matter} \bibinfo{volume}{\textbf{4}},
  \bibinfo{pages}{4251} (\bibinfo{date}{1992}).
\bibitem{Cin79}
\bibinfo{author}{M.~Cini}, \bibinfo{journal}{Surf. Sci.}
  \bibinfo{volume}{\textbf{87}}, \bibinfo{pages}{483} (\bibinfo{date}{1979}).
\bibitem{TDDS81}
\bibinfo{author}{G.~Tr{\'e}glia}, \bibinfo{author}{M.~C. Desjonqu{\`e}res},
  \bibinfo{author}{F.~Ducastelle}, and \bibinfo{author}{D.~Spanjaard},
  \bibinfo{journal}{J. Phys. C} \bibinfo{volume}{\textbf{14}},
  \bibinfo{pages}{4347} (\bibinfo{date}{1981}).
\bibitem{DK84}
\bibinfo{author}{V.~Drchal} and \bibinfo{author}{J.~Kudrnovsk{\'y}},
  \bibinfo{journal}{J. Phys. F} \bibinfo{volume}{\textbf{14}},
  \bibinfo{pages}{2443} (\bibinfo{date}{1984}).
\bibitem{Nol90}
\bibinfo{author}{W.~Nolting}, \bibinfo{journal}{Z. Phys. B}
  \bibinfo{volume}{\textbf{80}}, \bibinfo{pages}{73} (\bibinfo{date}{1990}).
\bibitem{NGE91}
\bibinfo{author}{W.~Nolting}, \bibinfo{author}{G.~Geipel}, and
  \bibinfo{author}{K.~Ertl}, \bibinfo{journal}{Phys. Rev. B}
  \bibinfo{volume}{\textbf{44}}(22), \bibinfo{pages}{12197}
  (\bibinfo{date}{1991}).
\bibitem{PBBN93}
\bibinfo{author}{M.~Potthoff}, \bibinfo{author}{J.~Braun},
  \bibinfo{author}{G.~Borstel}, and \bibinfo{author}{W.~Nolting},
  \bibinfo{journal}{Phys. Rev. B} \bibinfo{volume}{\textbf{74}}(19),
  \bibinfo{pages}{12480} (\bibinfo{date}{1993}).
\bibitem{PBB94}
\bibinfo{author}{M.~Potthoff}, \bibinfo{author}{J.~Braun}, and
  \bibinfo{author}{G.~Borstel}, \bibinfo{journal}{Z. Phys. B}
  \bibinfo{volume}{\textbf{95}}, \bibinfo{pages}{207} (\bibinfo{date}{1994}).
\bibitem{PBNB94}
\bibinfo{author}{M.~Potthoff}, \bibinfo{author}{J.~Braun},
  \bibinfo{author}{W.~Nolting}, and \bibinfo{author}{G.~Borstel},
  \bibinfo{journal}{Surf. Sci.} \bibinfo{volume}{\textbf{307-309}},
  \bibinfo{pages}{942} (\bibinfo{date}{1994}).
\bibitem{PBBN95}
\bibinfo{author}{M.~Potthoff}, \bibinfo{author}{J.~Braun},
  \bibinfo{author}{G.~Borstel}, and \bibinfo{author}{W.~Nolting},
  \bibinfo{journal}{J. Electron Spectrosc. Rel. Phen.}
  \bibinfo{volume}{\textbf{72}}, \bibinfo{pages}{163} (\bibinfo{date}{1995}).
\bibitem{AGD}
\bibinfo{author}{A.~A. Abrikosov}, \bibinfo{author}{L.~P. Gorkov}, and
  \bibinfo{author}{I.~E. Dzyaloshinski}, \bibinfo{title}{\emph{Methods of
  Quantum Field Theory in Statistical Mechanics}} (\bibinfo{publisher}{Dover},
  New York, \bibinfo{year}{1975}).
\bibitem{KM81}
\bibinfo{author}{L.~Kleinman} and \bibinfo{author}{K.~Mednick},
  \bibinfo{journal}{Phys. Rev. B} \bibinfo{volume}{\textbf{24}}(12),
  \bibinfo{pages}{6880} (\bibinfo{date}{1981}).
\bibitem{SAS92}
\bibinfo{author}{M.~M. Steiner}, \bibinfo{author}{R.~C. Albers}, and
  \bibinfo{author}{L.~J. Sham}, \bibinfo{journal}{Phys. Rev. B}
  \bibinfo{volume}{\textbf{45}}(23), \bibinfo{pages}{13272}
  (\bibinfo{date}{1992}).
\bibitem{Pap86}
\bibinfo{author}{D.~A. Papaconstantopoulos}, \bibinfo{title}{\emph{Handbook of
  the band structure of elemental solids}} (\bibinfo{publisher}{Plenum}, New
  York, \bibinfo{year}{1986}).
\bibitem{STK70}
\bibinfo{author}{S.~Sugano}, \bibinfo{author}{Y.~Tanabe}, and
  \bibinfo{author}{H.~Kamimura}, \bibinfo{title}{\emph{Multiplets of
  transition-metal ions in crystals}}, \bibinfo{volume}{vol.~33 of \emph{Pure
  and applied physics}} (\bibinfo{publisher}{Academic}, New York,
  \bibinfo{year}{1970}).
\bibitem{AAL97}
\bibinfo{author}{V.~I. Anisimov}, \bibinfo{author}{F.~Aryasetiawan}, and
  \bibinfo{author}{A.~I. Lichtenstein}, \bibinfo{journal}{J. Phys.: Condens.
  Matter} \bibinfo{volume}{\textbf{9}}, \bibinfo{pages}{767}
  (\bibinfo{date}{1997}).
\bibitem{Zub60}
\bibinfo{author}{D.~N. Zubarev}, \bibinfo{journal}{Sov. Phys. Uspekhi}
  \bibinfo{volume}{\textbf{3}}, \bibinfo{pages}{320} (\bibinfo{date}{1960}).
\bibitem{Nol7}
\bibinfo{author}{W.~Nolting}, \bibinfo{title}{\emph{Vielteilchentheorie}},
  \bibinfo{volume}{vol.~7 of \emph{Grundkurs: Theoretische Physik}}
  (\bibinfo{publisher}{Zimmermann-Neufang, Ulmen}, \bibinfo{year}{1995}).
\bibitem{SC91}
\bibinfo{author}{H.~Schweitzer} and \bibinfo{author}{G.~Czycholl},
  \bibinfo{journal}{Z. Phys. B} \bibinfo{volume}{\textbf{83}},
  \bibinfo{pages}{93} (\bibinfo{date}{1991}).
\bibitem{PN97}
\bibinfo{author}{M.~Potthoff} and \bibinfo{author}{W.~Nolting},
  \bibinfo{journal}{Z. Phys. B} \bibinfo{volume}{\textbf{104}},
  \bibinfo{pages}{265} (\bibinfo{date}{1997}).
\bibitem{LV84}
\bibinfo{author}{P.~Lambin} and \bibinfo{author}{J.~P. Vigneron},
  \bibinfo{journal}{Phys. Rev. B} \bibinfo{volume}{\textbf{29}}(6),
  \bibinfo{pages}{3430} (\bibinfo{date}{1984}).
\bibitem{APKAK97}
\bibinfo{author}{V.~I. Anisimov}, \bibinfo{author}{A.~I. Poteryaev},
  \bibinfo{author}{M.~A. Korotin}, \bibinfo{author}{A.~O. Anokhin}, and
  \bibinfo{author}{G.~Kotliar}, \bibinfo{journal}{J. Phys.: Condens. Matter}
  \bibinfo{volume}{\textbf{9}}, \bibinfo{pages}{7359} (\bibinfo{date}{1997}).
\bibitem{HS73}
\bibinfo{author}{S.~Hirooka} and \bibinfo{author}{M.~Shimizu},
  \bibinfo{journal}{Phys. Lett.} \bibinfo{volume}{\textbf{46A}},
  \bibinfo{pages}{209} (\bibinfo{date}{1973}).
\bibitem{FOH97}
\bibinfo{author}{M.~Fleck}, \bibinfo{author}{A.~M. Ole\'{s}}, and
  \bibinfo{author}{L.~Hedin}, \bibinfo{journal}{Phys. Rev. B}
  \bibinfo{volume}{\textbf{56}}(6), \bibinfo{pages}{3159}
  (\bibinfo{date}{1997}).
\bibitem{GBP77}
\bibinfo{author}{C.~Guilllot}, \bibinfo{author}{Y.~Ballu},
  \bibinfo{author}{J.~Paigne}, \bibinfo{author}{J.~Lecante},
  \bibinfo{author}{K.~P. Jain}, \bibinfo{author}{P.~Thiry},
  \bibinfo{author}{R.~Pinchaux}, \bibinfo{author}{Y.~Petroff}, and
  \bibinfo{author}{L.~M. Falicov}, \bibinfo{journal}{Phys. Rev. Lett.}
  \bibinfo{volume}{\textbf{39}}, \bibinfo{pages}{1632} (\bibinfo{date}{1977}).
\bibitem{SKO87}
\bibinfo{author}{Y.~Sakisaki}, \bibinfo{author}{T.~Komeda},
  \bibinfo{author}{M.~Ouchi}, \bibinfo{author}{H.~Kato},
  \bibinfo{author}{S.~Masuda}, and \bibinfo{author}{K.~Yagi},
  \bibinfo{journal}{Phys. Rev. Lett.} \bibinfo{volume}{\textbf{58}},
  \bibinfo{pages}{733} (\bibinfo{date}{1987}).
\bibitem{RM87}
\bibinfo{author}{S.~Raaen} and \bibinfo{author}{V.~Murgai},
  \bibinfo{journal}{Phys. Rev. B} \bibinfo{volume}{\textbf{36}},
  \bibinfo{pages}{887} (\bibinfo{date}{1987}).
\bibitem{Lie81}
\bibinfo{author}{A.~Liebsch}, \bibinfo{journal}{Phys. Rev. B}
  \bibinfo{volume}{\textbf{23}}(10), \bibinfo{pages}{5203}
  (\bibinfo{date}{1981}).
\bibitem{DJK98}
\bibinfo{author}{V.~Drchal}, \bibinfo{author}{V.~Jani{\v{s}}}, and
  \bibinfo{author}{J.~Kudranovsk{\'y}}, \bibinfo{journal}{cond-mat/9810181}
  (\bibinfo{date}{1998}), \url{http://xxx.lan.gov/abs/}.
\bibitem{BDZ91}
\bibinfo{author}{P.~J. Brown}, \bibinfo{author}{J.~Deportes}, and
  \bibinfo{author}{K.~R.~A. Ziebeck}, \bibinfo{journal}{J. Phys. I France}
  \bibinfo{volume}{\textbf{1}}, \bibinfo{pages}{1529} (\bibinfo{date}{1991}).
\bibitem{BDNZ92}
\bibinfo{author}{P.~J. Brown}, \bibinfo{author}{J.~Deportes},
  \bibinfo{author}{K.~U. Neumann}, and \bibinfo{author}{K.~R.~A. Ziebeck},
  \bibinfo{journal}{J. Magn. Magn. Mat.} \bibinfo{volume}{\textbf{104-107}},
  \bibinfo{pages}{2083} (\bibinfo{date}{1992}).
\bibitem{LB19a}
\bibinfo{author}{M.~B. Stearns}, \bibinfo{title}{\emph{Landoldt-B{\"o}rnstein,
  New Series}} (\bibinfo{publisher}{Springer}, Berlin, \bibinfo{year}{1984}),
  \bibinfo{volume}{vol. 19a of \emph{Group III}}, chap. Magnetic Properties of
  Metals.
\bibitem{AMTL87}
\bibinfo{author}{R.~Allenspach}, \bibinfo{author}{D.~Mauri},
  \bibinfo{author}{M.~Taborelli}, and \bibinfo{author}{M.~Landolt},
  \bibinfo{journal}{Phys. Rev. B} \bibinfo{volume}{\textbf{35}}(10),
  \bibinfo{pages}{4801} (\bibinfo{date}{1987}).
\bibitem{SK54}
\bibinfo{author}{J.~C. Slater} and \bibinfo{author}{G.~F. Koster},
  \bibinfo{journal}{Phys. Rev.} \bibinfo{volume}{\textbf{94}}(6),
  \bibinfo{pages}{1498} (\bibinfo{date}{1954}).
\bibitem{Loe50}
\bibinfo{author}{P.-O. L{\"o}wdin}, \bibinfo{journal}{J. Chem. Phys.}
  \bibinfo{volume}{\textbf{18}}(3), \bibinfo{pages}{365}
  (\bibinfo{date}{1950}).

\end{thebibliography}
\end{document}